\documentclass[11pt,superscriptaddress,aps]{revtex4}
\usepackage{graphicx}

\begin{document}
\title{SUM RULES FOR TOTAL CROSS-SECTIONS OF HADRON PHOTO-PRODUCTION ON
       PSEUDOSCALAR MESONS AND OCTET BARYONS}
\date{\today}

\author{E.Barto\v s}\address{Institute of Physics, Slovak Academy of Sciences,
Bratislava, Slovak Republic}

\author{\underline{S.Dubni\v cka}}
\address{Institute of Physics, Slovak Academy of Sciences,
Bratislava, Slovak Republic}

\author{A.Z.Dubni\v ckov\'a}
\address{Dept. of Theoretical Physics, Comenius Univ., Bratislava,
Slovak Republic}

\author{E.A.Kuraev}
\address{Laboratory of Theoretical Physics, JINR, Dubna, Russia}

\vspace{2cm}
\begin{abstract}
Considering very high energy peripheral electron-hadron scattering
with a production of hadronic state $X$ moving closely to the
direction of initial hadron the Weizs\"acker-Williams like
expression, relating the difference of $q^2$-dependent
differential cross-sections of DIS processes to the convergent
integral over the difference of the total hadron photo-production
cross-sections on hadros, is derived. Then, exploiting analytic
properties of the forward retarded Compton scattering amplitude on
the same hadrons, first, the sum rules are derived bringing into
relation hadron electromagnetic form factors with the difference
of the $q^2$-dependent differential cross-sections of DIS, then
using Weizs\"acker-Williams like expression and taking the
derivative of both sides according to ${\bf q^2}$ for ${\bf q^2}
\to 0$ one comes to new universal hadron sum rules relating hadron
static parameters to the convergent integral over the difference
of the total hadron photo-production cross-sections on hadrons.

\end{abstract}

\maketitle

\section{WEIZS{\"A}CKER-WILLIAMS LIKE RELATIONS}

 In a derivation of them one considers a very high energy peripheral electro-production
process on hadrons $h$
\begin{equation}
e^-(p_1) + h(p) \to e^-(p_1') + X, \label{a1}
\end{equation}
with the produced pure hadronic state $X$  moving closely to the
direction of the initial hadron to be described by the matrix
element
\begin{equation}
M = i \frac{\sqrt{4\pi \alpha}}{q^2} \bar u(p_1^{'})\gamma_\mu
u(p_1) <X \mid J_\nu^{EM} \mid h>g^{\mu\nu},\label{a2}
\end{equation}
in the one photon exchange approximation with  $m^2_X = (p+q)^2$ .

   Now, applying the Sudakov expansion of the photon transferred four-vector
$q$
\begin{equation}\label{a3}
q=\beta_q \tilde{p}_1+\alpha_q\tilde{p}+q^{\bot}, \quad
q_{\bot}=(0,0,{\bf q}),\quad q_{\bot}^2=-{\bf q}^2
\end{equation}
into the almost light-like vectors
\begin{equation}\label{a4}
\tilde{p}_1=p_1-m_e^2p/(2p_1p), \quad
\tilde{p}=p-m_B^2p_1/(2p_1p)
\end{equation}
and also the Gribov prescription for the numerator of the photon
Green function
\begin{equation}
g_{\mu\nu}=g_{\mu\nu}^{\bot}+\frac{2}{s}(\tilde{p}_{\mu}\tilde{p}_{1\nu}+\tilde{p}_{\nu}\tilde{p}_{1\mu})
 \approx\frac{2}{s} \tilde{p}_\mu\tilde{p}_{1\nu} \label{a5}
 \end{equation}
with $s=(p_1+p)^2\approx 2p_1p\gg Q^2 = -q^2$   in (\ref{a2}),
then for very high electron energy in (\ref{a1}) and small photon
momentum transfer squared $t=q^2=-Q^2=-{\bf q^2}$ the
cross-section can be written in the form
\begin{eqnarray}\label{a6}
&&d\sigma^{e^-h\to e^-X}= \frac{4\pi\alpha}{s(q^2)^2}
p_1^{\mu}p_1^{\nu}\times \\ &&\sum_{X\neq h}\sum_{r=-j}^{j}
\langle h^{(r)}\mid J_\mu^{EM}\mid X\rangle^* \langle X \mid
J_\nu^{EM}\mid h^{(r)}\rangle d \Gamma \nonumber
\end{eqnarray}
with a summation through the created hadronic states $X$ and the
spin states of the initial hadron.

If the relation $\int d^{4}q\delta^{(4)}(p_1-q-p_1')=1$ is used in
the phase space volume $d\Gamma$ of the final particles, then
\begin{equation}\label{a7}
d\Gamma = \frac{d s_1}{2s(2\pi)^3}d^2{\bf {q}}d\Gamma_X
\end{equation}
with
\begin{equation}
d\Gamma_X=(2\pi)^4
\delta^{(4)}(p+q-\sum_i^nq_i)\prod_i^n\frac{d^3q_i}{2E_i(2\pi)^3}\label{a8}
\end{equation}
\begin{equation}
 s_1=2(qp)=m_X^2+{\bf q^2}-m^2_{h}=s\beta_q.
 \label{a9}
 \end{equation}

Moreover, the current conservation condition ($\alpha_q\tilde p$
gives a negligible contribution)
\begin{eqnarray}\label{a10}
&&q^\mu \langle X\mid J_\mu^{EM}\mid h^{(r)}\rangle \approx \\
\nonumber&&\approx(\beta_q\tilde{p}_1+q_\bot)^\mu\langle X\mid
J_\mu^{EM}\mid h^{(r)}\rangle = 0,
\end{eqnarray}
is applied in order to utilize in the expression for cross-section
the relation
\begin{eqnarray}\nonumber
\int p_1^\mu p_1^\nu \sum_{X\neq h}\sum_{r=-j}^{j}\langle
h^{(r)}\mid J_\mu^{EM}\mid X\rangle^* &&\\ \langle X \mid
J_\nu^{EM}\mid h^{(r)}\rangle d \Gamma_X=
 \label{a11} 2i\frac{s^2}{s_1^2}{\bf {q}^2} Im \tilde {A}^{(h)}(s_1,{\bf
 {q}}),&&
\end{eqnarray}
with the imaginary part of the retarded forward Compton scattering
amplitude $\tilde {A}^h(s_1, \bf q)$  on a hadron.

As a result for a difference of corresponding differential
cross-sections of the electro-production on $h$ and $h'$ (after
integration in the cross-section (\ref{a6}) over $d\Gamma_X$, as
well as over the invariant mass squared $m_X^2$, i.e. over the
variable $s_1$ to be interested only for ${\bf q}$ distribution)
one finds the expression
\begin{eqnarray} \nonumber
&&\Big(\frac{d\sigma^{e^-h\to e^-X}(s,{\bf q})}{d^2{\bf{q}}} -
\frac{d\sigma^{e^-h'\to
e^-X'}(s,{\bf q})}{d^2{\bf{q}}}\Big)=\\
\label{a12} &=&\frac{\alpha{\bf{q}^2}}
{4\pi^2}\int\limits_{s_1^{thr}}^\infty\frac{d
s_1}{s_1^2[{\bf{q}^2}+(m_es_1/s)^2]^2}\times \\
\nonumber&\times& [Im \tilde{A^h}(s_1,{\bf{q}})- Im
\tilde{A^{h'}}(s_1,{\bf{q}})].
\end{eqnarray}

Finally, if one neglects the second term in square brackets of the
denominator of the integral in (\ref{a12}) (as $m_e$ is  small and
 $s$ is large in comparison with $s_1$),
the expressions $d^2{\bf q}=\pi d{\bf{q}}^2$ is exploited, the
limit ${\bf{q}^2}\to 0$ along with the relation $Im
\tilde{A}^h(s_1,{\bf{q}})$= $4s_1\sigma_{tot}^{\gamma^{*}h\to
X}(s_1,{\bf{q}})$ is applied, one comes to Weizs{\"a}cker-Williams
like expressions
\begin{eqnarray} \label{a13}
&&{\bf{q}}^2\Big(\frac{d\sigma^{e^- h\to e^-X}}{d{\bf{q}}^2} -
\frac{d\sigma^{e^- h'\to e^-X'}}{d{\bf{q}^2}}\Big)_
{|_{{\bf{q}^2}\to 0}}=\\
\nonumber &=&\frac{\alpha}
{\pi}\int\limits_{s_1^{thr}}^\infty\frac{d s_1}{s_1}
[\sigma_{tot}^{\gamma h\to X}(s_1)-\sigma_{tot}^{\gamma h'\to
X}(s_1)],
\end{eqnarray}
relating the difference of $q^2$-dependent differential
cross-sections of the DIS processes to the convergent integral
over the difference of the total hadron photo-production
cross-sections on hadrons.

From the relation (\ref{a13}) one can see immediately that
considering differences of the total cross-sections one achieves
the convergent integral.

\section{VIRTUAL COMPTON SCATTERING ON HADRON}

In the previous section we have used the concept "the retarded
forward Compton scattering amplitude on hadron". Here we slightly
clarify it.

The total virtual photon Compton scattering amplitude
$A(s_1,\bf{q})$ consists of the four different contributions to be
classified according to the corresponding Feynman diagrams
\begin{equation}
A(s_1,{\bf q})=\tilde A(s_1,{\bf q})+ A_a(s_1,{\bf q})+
A_P(s_1,{\bf q})+A_{odd}. \label{eq:amp}
\end{equation}

$\tilde A(s_1,\bf{q})$ represents a class of diagrams in which the
initial state photon is first absorbed by a hadron line and then
emitted by the scattered hadron - retarded Compton scattering
amplitude

$A_a(s_1,\bf{q})$ represents a class of diagrams in which the
scattered photon is first emitted along the hadron line and the
point of absorption is located later on - advanced Compton
scattering amplitude

$A_P(s_1,\bf{q})$ represents a class of diagrams in which both
photons do not interact with the initial hadron line, in other
words it corresponds to the Pomeron - type Feynman diagrams and
gives the non-vanishing contributions to the total cross-sections
in the limit of a large invariant mass squared of initial
particles $s_1\rightarrow \infty$

$A_{odd}$ can be relevant in experiments  measuring charge-odd
effects

For more detail see Ref.\cite{Ku1}

\section{$q^2$ DEPENDENT MESON AND BARYON SUM RULES}

\begin{figure*}[hbt]
\begin{center}
\includegraphics[scale=.7]{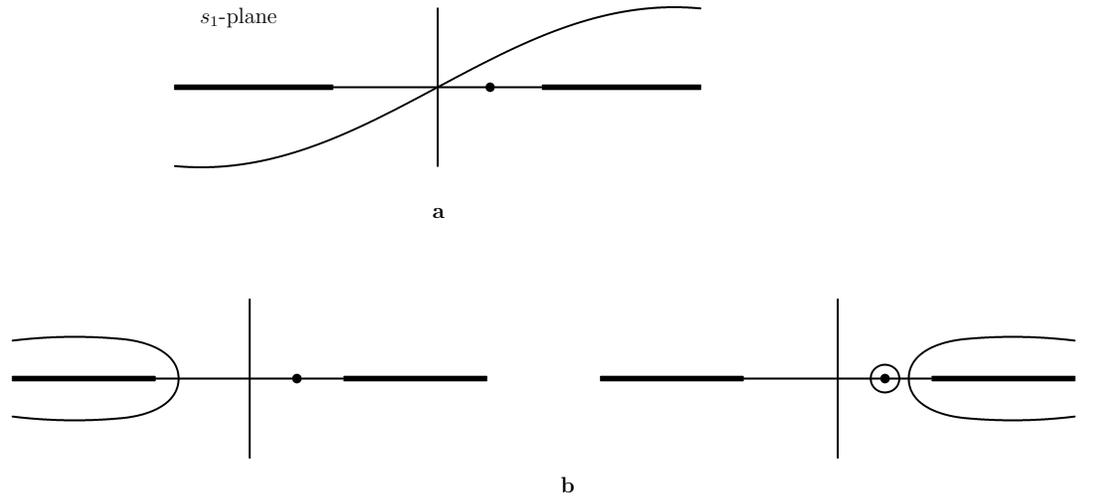}
\caption {\label{fig:1}Sum rule interpretation in $s_1$ plane.}
\end{center}
\end{figure*}

Universal sum rules are derived by investigating the analytic
properties of the retarded Compton scattering amplitude
$\tilde{A}^h(s_1,{\bf{q}})$ in $s_1$- plane as presented in Fig.
1a, then defining the integral $I$ over the path $C$ (for more
detail see \cite{Bai}) in the $s_1$-plane
\begin{equation}\label{a14}
I=\int\limits_C d s_1 \frac{p_1^\mu p_1^\nu}{s^2}
\left(\tilde{A}^{h}_{\mu\nu}(s_1,{\bf{q}})-
\tilde{A}^{h'}_{\mu\nu}(s_1,{\bf{q}})\right )
\end{equation}
from the gauge invariant light-cone projection
$p_1^{\mu}p_1^{\nu}\tilde{A}_{\mu\nu}^h(s_1,{\bf{q}})$ of the
amplitude $\tilde{A}^h(s_1,{\bf{q}})$ and once closing the contour
$C$ to upper half-plane, another one to lower half-plane (see Fig.
1b).

As a result the following sum rule appears
\begin{eqnarray}\nonumber
& &\pi (Res^{h'}-Res^{h})
={\bf{q}}^2\int\limits_{r.h.}^\infty\frac{ds_1}
{s_1^2} [Im \tilde{A}^{h}(s_1,{\bf{q}})-\\
&-& Im \tilde{A}^{h'}(s_1,{\bf{q}})]. \label{a15}
\end{eqnarray}

The left-hand cut contributions expressed by an integral over the
difference $[Im \tilde{A}^{h}(s_1,{\bf{q}}) - Im
\tilde{A}^{h'}(s_1,{\bf{q}})]$ are assumed to be mutually
annulated.

Now, if the meson or baryon sum rules are considered, one has to
take into account the corresponding residuum of the intermediate
state pole (see Fig. 1).

If one considers mesons - their electromagnetic structure is
described by one charge form factor and the residuum takes the
form

\begin{equation}
\label{a16} Res^{(M)}=2\pi \alpha F^2_M(\bf{-q^2}).
\end{equation}

If one considers baryons - their electromagnetic structure is
described by Dirac and Pauli form factors and the residuum takes
different form
\begin{equation}\label{a17}
Res^{B}=2\pi\alpha( F^2_{1B}+ \frac{{\bf{q}^2}}{4m_B^2}F_{2B}^2),
\end{equation}
where in both cases an averaging over the initial hadron and
photon spins is performed.

Then, substituting (\ref{a16}) and (\ref{a17}) into (\ref{a15})
and taking into account (\ref{a12}) from the previous Section with
$d^2{\bf q}=\pi d{\bf q^2}$, one comes to the $q^2$- dependent
meson sum rule \cite{Du1}
\begin{eqnarray}
& &[F^2_{P'}({\bf{-q^2}})-F^2_{P'}(0)] -
[F^2_P({\bf{-q^2}})-F^2_P(0)] = \nonumber\\
&=&\frac{2}{\pi \alpha^2}({\bf{q^2}})^2\Big(\frac{d\sigma^{e^-
P\to e^-X}}{d{\bf{q}}^2} - \frac{d\sigma^{e^- P'\to
e^-X'}}{d{\bf{q}}^2}\Big)\label{a18},
\end{eqnarray}
and the $q^2$-dependent baryon sum rule \cite{Du2}
\begin{eqnarray}
& &[F^2_{1B'}({-\bf{q^2}})-F^2_{1B'}(0)] -
[F^2_{1B}({\bf{-q^2}})-F^2_{1B}(0)] + \nonumber\\
&+&{\bf{q^2}}\big [\frac{F^2_{2B'}({\bf-{q^2}})}{4m_{B'}^2}-
\frac{F^2_{2B}({\bf-{q^2}})}{4m_{B}^2}\big ]=\nonumber \\
 &=&\frac{2}{\pi
\alpha^2}({\bf{q^2}})^2\Big(\frac{d\sigma^{e^- B\to
e^-X}}{d{\bf{q}}^2} - \frac{d\sigma^{e^- B'\to
e^-X}}{d{\bf{q}}^2}\Big ),\label{a19}
\end{eqnarray}
respectively, where the left-hand side in both cases was
re-normalized in order to separate the pure strong interactions
from electromagnetic ones.

\section{UNIVERSAL SUM RULE FOR TOTAL HADRON PHOTO-PRODUCTION
CROSS-SECTIONS ON MESONS}

Now, employing the Weics\"acker-Williams like relation for mesons,
taking a derivative according to ${\bf q^2}$ of both sides in
$q^2$-dependent meson sum rule for ${\bf q^2} \to 0$ and using the
laboratory reference frame by $s_1=2m_B\omega$, one comes
 to the new universal meson sum rule \cite{Du1} relating meson mean
square radii to the integral over a difference of the
corresponding total photo-production cross-sections on mesons
\begin{eqnarray}\label{a20}
&&\frac{1}{3}(\langle r_{P'}^2 \rangle-\langle r_P^2 \rangle)=\\
&=&\frac{2}{\pi^2\alpha}\int\limits_{\omega_P}^\infty
\frac{d\omega} {\omega}\big[\sigma_{tot}^{\gamma P\to X}(\omega)-
\sigma_{tot}^{\gamma P'\to X}(\omega)\big], \nonumber
\end{eqnarray}
in which just a mutual cancellation of the rise of the latter
cross sections for $\omega \to \infty$ is achieved.

\section{APPLICATION TO VARIOUS COUPLES OF MESONS}

According to the SU(3) classification of existing hadrons the
following ground state pseudoscalar meson nonet $\pi^-$, $\pi^0$,
$\pi^+$, $K^-$, $\bar K^0$, $K^0$, $K^+$, $\eta$, $\eta'$ exists.
However, in consequence of CPT invariance the meson
electromagnetic form factors $F_P(-{\bf q^2})$ hold the following
relation
\begin{equation}
F_P({\bf -q^2})= - F_{\bar P}({\bf -q^2}),\label{a21}
\end{equation}
where $\bar P$ means antiparticle.

Since $\pi_0$, $\eta$ and $\eta'$ are true neutral particles,
their electromagnetic form factors are owing to the (\ref{a21})
zero in the whole region of a definition and therefore we exclude
them from further considerations.

If one considers couples of particle-antiparticle like
$\pi^{\pm}$, $K^{\pm}$ and $K^0$, $\bar K^0$, the left hand side
of (\ref{a18}) is owing to the relation (\ref{a21}) equal zero and
we exclude couples $\pi^{\pm}$, $K^{\pm}$ and $K^0$, $\bar K^0$
from further considerations as well.

If one  considers a couple of the iso-doublet of kaons $K^+, K^0$
and $K^-, \bar K^0$, the following Cabibbo-Radicati like sum rules
\cite{Cab} for kaons can be written
\begin{eqnarray}\label{a22}
\frac{1}{6}{\pi^2\alpha}\langle r_{K^+}^2\rangle
=\int_{\omega_{th}}^{\infty} \frac{d\omega}
{\omega}\left[\sigma_{tot}^{\gamma K^+\to X}(\omega)-
\sigma_{tot}^{\gamma K^0\to X}(\omega)\right]
\end{eqnarray}

\begin{eqnarray}\label{a23}
&&\frac{1}{6}{\pi^2\alpha}(-1)\langle r_{K^-}^2\rangle=
\int_{\omega_{th}}^{\infty} \frac{d\omega}
{\omega}\left[\sigma_{tot}^{\gamma K^-\to X}(\omega)-
\sigma_{tot}^{\gamma \bar K^0\to X}(\omega)\right],
\end{eqnarray}
in which the relation $\langle r_{K^+}^2\rangle=-\langle
r_{K^-}^2\rangle$ for kaon mean squared charge radii, following
directly from (\ref{a21}), holds and divergence of the integrals,
due to an increase of the total cross-sections
$\sigma_{tot}^{\gamma K^\pm\to X}(\omega)$ for large values of
$\omega$, is taken off by the increase of total cross-sections
$\sigma_{tot}^{\gamma K^0\to X}(\omega)$ and $\sigma_{tot}^{\gamma
\bar K^0\to X}(\omega)$, respectively.
   If besides the latter, also the relations
\begin{eqnarray}\label{a24}
\sigma_{tot}^{\gamma K^0\to X}(\omega)\equiv\sigma_{tot}^{\gamma
\bar K^0\to X}(\omega)\\\nonumber \sigma_{tot}^{\gamma K^+\to
X}(\omega)\equiv\sigma_{tot}^{\gamma K^-\to X}(\omega),
\end{eqnarray}
following from $C$ invariance of the electromagnetic interactions,
are taken into account, one can see the sum rule (\ref{a23}), as
well as all other possible sum rules obtained by combinations
$K^+\bar K^0, K^-K^0,$ to be contained already in (\ref{a22}).

The last possibility is a consideration of a couple of mesons
taken from the isomultiplet of pions and the isomultiplet of kaons
leading to the following sum rules
\begin{eqnarray}\label{a25}
&&\frac{1}{6}{\pi^2\alpha}[(\pm1)\langle r_{\pi^{\pm}}^2\rangle-
(\pm1)\langle r_{K^{\pm}}^2\rangle]= int_{\omega_{th}}^{\infty}
\frac{d\omega} {\omega}\left[\sigma_{tot}^{\gamma \pi^{\pm}\to
X}(\omega)- \sigma_{tot}^{\gamma K^{\pm}\to X}(\omega)\right]
\end{eqnarray}

\begin{eqnarray}\label{a26}
 &&\frac{1}{6}{\pi^2\alpha}(\pm1)\langle r_{\pi^{\pm}}^2\rangle =
\int_{\omega_{th}}^{\infty} \frac{d\omega}
{\omega}\left[\sigma_{tot}^{\gamma \pi^{\pm}\to X}(\omega)-
\sigma_{tot}^{\gamma K^0\to X}(\omega)\right].
\end{eqnarray}

   Now taking the experimental values \cite{Rev}\\
$(\pm1)\langle r_{\pi^{\pm}}^2\rangle=+0.4516\pm0.0108 \quad
[fm^2]$ \quad
$(\pm1)\langle r_{K^{\pm}}^2\rangle=+0.3136\pm0.0347 \quad [fm^2]$\\
one comes to the conclusion that in average

\begin{eqnarray}
[\sigma_{tot}^{\gamma \pi^{\pm}\to X}(\omega)-
\sigma_{tot}^{\gamma K^{\pm}\to X}(\omega)] > 0\\ \nonumber
[\sigma_{tot}^{\gamma K^-\to X}(\omega)- \sigma_{tot}^{\gamma \bar
K^0\to X}(\omega)] > 0,
\end{eqnarray}
from where the following chain of inequalities for finite values
of $\omega$ in average follow
\begin{eqnarray}
\sigma_{tot}^{\gamma \pi^{\pm}\to X}(\omega)> \sigma_{tot}^{\gamma
K^{\pm}\to X}(\omega) > \sigma_{tot}^{\gamma \bar K^0\to
X}(\omega) > 0.
\end{eqnarray}

Subtracting up (\ref{a22}) or (\ref{a23}) from the relation
(\ref{a26}), the sum rule (\ref{a25}) is obtained, what
demonstrates a mutual consistency of all considered sum rules.

They have been derived in analogy with a derivation  of the sum
rule for a difference of proton and neutron total photo-production
cross-sections \cite{Bar}, which is fulfilled with a very high
precision. Therefore we believe that also the sum rules for total
cross-sections of hadron photo-production on pseudoscalar mesons
presented in this paper are correct.

\section{UNIVERSAL SUM RULE FOR TOTAL HADRON PHOTO-PRODUCTION
CROSS-SECTIONS ON BARYONS}

Now employing the Weics\"acker-Williams relation for baryons,
taking a derivative  according to ${\bf{q}}^2$ of both sides in
$q^2$-dependent baryon sum rule for ${\bf{q}^2}\to 0$ and using
the laboratory reference frame by $s_1=2m_B\omega$ , one comes to
the new universal baryon sum rule \cite{Du2}
\begin{eqnarray}\nonumber
&&\frac{1}{3}\big [F_{1B}(0)\langle r_{1B}^2
\rangle-F_{1B'}(0)\langle r_{1B'}^2 \rangle\big ]-
\big [\frac{\kappa_B^2}{4m_B^2}-\frac{\kappa_{B'}^2}{4m^2_{B'}}\big ]=\\
\label{a27}
&=&\frac{2}{\pi^2\alpha}\int\limits_{{\omega_B}}^{\infty}
\frac{d\omega} {\omega}\big[\sigma_{tot}^{\gamma B\to X}(\omega)-
\sigma_{tot}^{\gamma B'\to X}(\omega)\big]
\end{eqnarray}
relating Dirac baryon mean square radii $\langle r_{1B}^2\rangle$
and baryon anomalous magnetic moments $\kappa_B$ to the convergent
integral, in which a mutual cancellation of the rise of the
corresponding total cross-sections for $\omega\to\infty$ is
achieved.

\section{APPLICATION TO VARIOUS COUPLES OF OCTET BARYONS}

According to the SU(3) classification of existing hadrons - there
are known the following members of the ground state $1/2^+$ baryon
octet ($p$, $n$, $\Lambda^0$, $\Sigma^+$,  $\Sigma^0$, $\Sigma^-$,
$\Xi^0$, $\Xi^-$). As a result, by using the universal expression
(\ref{a27}) one can write down $8!/(2!(8-2)!)=28$ different sum
rules for total cross-sections of hadron photo-production on
ground state $1/2^+$ octet baryons.

The most interesting from the point of view of experimental
verification is proton-neutron sum rule \cite{Bar}
\begin{equation}\nonumber
\frac{1}{3}\langle r_{1p}^2\rangle
-\frac{\kappa_{p}^2}{4m_{p}^2}+\frac{\kappa_{n}^2}{4m_{n}^2}=
\frac{2}{\pi^2\alpha}\int\limits_{\omega_{N}}^\infty
\frac{d\omega}{\omega} \big[\sigma_{tot}^{\gamma {p\to
X}}(\omega)- \sigma_{tot}^{\gamma {n\to X}}(\omega)\big].
\end{equation}
If one considers couples of the iso-triplet of $\Sigma$-hyperons
and separately couples of the iso-doublet of $\Xi$-hyperons, one
finds
\begin{eqnarray}\nonumber
&&\frac{1}{3}\big [\langle r_{1\Sigma^+}^2 \rangle -
\big [\frac{\kappa_{\Sigma^+}^2}{4m_{\Sigma^+}^2}-\frac{\kappa_{\Sigma^0}^2}{4m^2_{\Sigma^0}}\big ]=\\
\label{a28}
&=&\frac{2}{\pi^2\alpha}\int\limits_{\omega_{\Sigma^+}}^{\infty}
\frac{d\omega} {\omega}\big[\sigma_{tot}^{\gamma \Sigma^+\to
X}(\omega)- \sigma_{tot}^{\gamma \Sigma^0\to X}(\omega)\big ],
\end{eqnarray}
\begin{eqnarray}\nonumber
&&\frac{1}{3}\big [\langle r_{1\Sigma^+}^2 \rangle -\langle
r_{1\Sigma^-}^2 \rangle\big ]-
\big [\frac{\kappa_{\Sigma^+}^2}{4m_{\Sigma^+}^2}-\frac{\kappa_{\Sigma^-}^2}{4m^2_{\Sigma^-}}\big ]=\\
\label{a29}
&=&\frac{2}{\pi^2\alpha}\int\limits_{\omega_{\Sigma^+}}^{\infty}
\frac{d\omega} {\omega}\big[\sigma_{tot}^{\gamma \Sigma^+\to
X}(\omega)- \sigma_{tot}^{\gamma \Sigma^-\to X}(\omega)\big ],
\end{eqnarray}
\begin{eqnarray}\nonumber
&&\frac{1}{3}\langle r_{1\Sigma^-}^2 \rangle-
\big [\frac{\kappa_{\Sigma^0}^2}{4m_{\Sigma^0}^2}-\frac{\kappa_{\Sigma^-}^2}{4m^2_{\Sigma^-}}\big ]=\\
\label{a30}
&=&\frac{2}{\pi^2\alpha}\int\limits_{\omega_{\Sigma^0}}^{\infty}
\frac{d\omega} {\omega}\big[\sigma_{tot}^{\gamma \Sigma^0\to
X}(\omega)- \sigma_{tot}^{\gamma \Sigma^-\to X}(\omega)\big ],
\end{eqnarray}
and
\begin{eqnarray}\nonumber
&&\frac{1}{3}\langle r_{1\Xi^-}^2 \rangle-
\big [\frac{\kappa_{\Xi^0}^2}{4m_{\Xi^0}^2}-\frac{\kappa_{\Xi^-}^2}{4m^2_{\Xi^-}}\big ]=\\
\label{a31}
&=&\frac{2}{\pi^2\alpha}\int\limits_{\omega_{\Xi^0}}^{\infty}
\frac{d\omega} {\omega}\big[\sigma_{tot}^{\gamma \Xi^0\to
X}(\omega)- \sigma_{tot}^{\gamma \Xi^-\to X}(\omega)\big ],
\end{eqnarray}
respectively, which represent the second class of the baryon sum
rules.

The third class of the 23 baryon sum rules is found by a
consideration of a couple of baryons always taken from different
isomultiplets of the ground state $1/2^+$ baryon octet and take
forms as follows
\begin{eqnarray}\nonumber
&&\frac{1}{3}\langle r_{1p}^2 \rangle-
\big [\frac{\kappa_{p}^2}{4m_{p}^2}-\frac{\kappa_{\Lambda^0}^2}{4m^2_{\Lambda^0}}\big ]=\\
\label{a32}
&=&\frac{2}{\pi^2\alpha}\int\limits_{\omega_{p}}^{\infty}
\frac{d\omega} {\omega}\big[\sigma_{tot}^{\gamma p\to X}(\omega)-
\sigma_{tot}^{\gamma \Lambda^0\to X}(\omega)\big ],
\end{eqnarray}

\begin{eqnarray}\nonumber
&&\frac{1}{3}\big [\langle r_{1p}^2 \rangle -\langle r_{1\Sigma^+}^2
\rangle\big ]-
\big [\frac{\kappa_{p}^2}{4m_{p}^2}-\frac{\kappa_{\Sigma^+}^2}{4m^2_{\Sigma^+}}\big ]=\\
\label{a33}
&=&\frac{2}{\pi^2\alpha}\int\limits_{\omega_{p}}^{\infty}
\frac{d\omega} {\omega}\big[\sigma_{tot}^{\gamma p \to X}(\omega)-
\sigma_{tot}^{\gamma \Sigma^+\to X}(\omega)\big ],
\end{eqnarray}

\begin{eqnarray}\nonumber
&&\frac{1}{3}\langle r_{1p}^2 \rangle-
\big [\frac{\kappa_{p}^2}{4m_{p}^2}-\frac{\kappa_{\Sigma^0}^2}{4m^2_{\Sigma^0}}\big ]=\\
\label{a34}
&=&\frac{2}{\pi^2\alpha}\int\limits_{\omega_{p}}^{\infty}
\frac{d\omega} {\omega}\big[\sigma_{tot}^{\gamma p\to X}(\omega)-
\sigma_{tot}^{\gamma \Sigma^0\to X}(\omega)\big ],
\end{eqnarray}

\begin{eqnarray}\nonumber
&&\frac{1}{3}\big [\langle r_{1p}^2 \rangle +\langle r_{1\Sigma^-}^2
\rangle\big ]-
\big [\frac{\kappa_{p}^2}{4m_{p}^2}-\frac{\kappa_{\Sigma^-}^2}{4m^2_{\Sigma^-}}\big ]=\\
\label{a35}
&=&\frac{2}{\pi^2\alpha}\int\limits_{\omega_{p}}^{\infty}
\frac{d\omega} {\omega}\big[\sigma_{tot}^{\gamma p \to X}(\omega)-
\sigma_{tot}^{\gamma \Sigma^-\to X}(\omega)\big ],
\end{eqnarray}

\begin{eqnarray}\nonumber
&&\frac{1}{3}\langle r_{1p}^2 \rangle-
\big [\frac{\kappa_{p}^2}{4m_{p}^2}-\frac{\kappa_{\Xi^0}^2}{4m^2_{\Xi^0}}\big ]=\\
\label{a36}
&=&\frac{2}{\pi^2\alpha}\int\limits_{\omega_{p}}^{\infty}
\frac{d\omega} {\omega}\big[\sigma_{tot}^{\gamma p\to X}(\omega)-
\sigma_{tot}^{\gamma \Xi^0\to X}(\omega)\big ],
\end{eqnarray}

\begin{eqnarray}\nonumber
&&\frac{1}{3}\big [\langle r_{1p}^2 \rangle +\langle r_{1\Xi^-}^2
\rangle\big ]-
\big [\frac{\kappa_{p}^2}{4m_{p}^2}-\frac{\kappa_{\Xi^-}^2}{4m^2_{\Xi^-}}\big ]=\\
\label{a37}
&=&\frac{2}{\pi^2\alpha}\int\limits_{\omega_{p}}^{\infty}
\frac{d\omega} {\omega}\big[\sigma_{tot}^{\gamma p \to X}(\omega)-
\sigma_{tot}^{\gamma \Xi^-\to X}(\omega)\big ],
\end{eqnarray}

\begin{eqnarray}\nonumber
&&-\big [\frac{\kappa_{n}^2}{4m_{n}^2}-\frac{\kappa_{\Lambda^0}^2}{4m^2_{\Lambda^0}}\big ]=\\
\label{a38}
&=&\frac{2}{\pi^2\alpha}\int\limits_{\omega_{n}}^{\infty}
\frac{d\omega} {\omega}\big[\sigma_{tot}^{\gamma n\to X}(\omega)-
\sigma_{tot}^{\gamma \Lambda^0\to X}(\omega)\big ],
\end{eqnarray}

\begin{eqnarray}\nonumber
&&-\frac{1}{3}\langle r_{1\Sigma^+}^2 \rangle-
\big [\frac{\kappa_{n}^2}{4m_{n}^2}-\frac{\kappa_{\Sigma^+}^2}{4m^2_{\Sigma^+}}\big ]=\\
\label{a39}
&=&\frac{2}{\pi^2\alpha}\int\limits_{\omega_{n}}^{\infty}
\frac{d\omega} {\omega}\big[\sigma_{tot}^{\gamma n\to X}(\omega)-
\sigma_{tot}^{\gamma \Sigma^+\to X}(\omega)\big ],
\end{eqnarray}

\begin{eqnarray}\nonumber
&&-\big [\frac{\kappa_{n}^2}{4m_{n}^2}-\frac{\kappa_{\Sigma^0}^2}{4m^2_{\Sigma^0}}\big ]=\\
\label{a40}
&=&\frac{2}{\pi^2\alpha}\int\limits_{\omega_{n}}^{\infty}
\frac{d\omega} {\omega}\big[\sigma_{tot}^{\gamma n\to X}(\omega)-
\sigma_{tot}^{\gamma \Sigma^0\to X}(\omega)\big ],
\end{eqnarray}

\begin{eqnarray}\nonumber
&&\frac{1}{3}\langle r_{1\Sigma^-}^2 \rangle-
\big [\frac{\kappa_{n}^2}{4m_{n}^2}-\frac{\kappa_{\Sigma^-}^2}{4m^2_{\Sigma^-}}\big ]=\\
\label{a41}
&=&\frac{2}{\pi^2\alpha}\int\limits_{\omega_{n}}^{\infty}
\frac{d\omega} {\omega}\big[\sigma_{tot}^{\gamma n\to X}(\omega)-
\sigma_{tot}^{\gamma \Sigma^-\to X}(\omega)\big ],
\end{eqnarray}

\begin{eqnarray}\nonumber
&&-\big [\frac{\kappa_{n}^2}{4m_{n}^2}-\frac{\kappa_{\Xi^0}^2}{4m^2_{\Xi^0}}\big ]=\\
\label{a42}
&=&\frac{2}{\pi^2\alpha}\int\limits_{\omega_{n}}^{\infty}
\frac{d\omega} {\omega}\big[\sigma_{tot}^{\gamma n\to X}(\omega)-
\sigma_{tot}^{\gamma \Xi^0\to X}(\omega)\big ],
\end{eqnarray}

\begin{eqnarray}\nonumber
&&\frac{1}{3}\langle r_{1\Xi^-}^2 \rangle-
\big [\frac{\kappa_{n}^2}{4m_{n}^2}-\frac{\kappa_{\Xi^-}^2}{4m^2_{\Xi^-}}\big ]=\\
\label{a43}
&=&\frac{2}{\pi^2\alpha}\int\limits_{\omega_{n}}^{\infty}
\frac{d\omega} {\omega}\big[\sigma_{tot}^{\gamma n\to X}(\omega)-
\sigma_{tot}^{\gamma \Xi^-\to X}(\omega)\big ],
\end{eqnarray}

\begin{eqnarray}\nonumber
&&-\frac{1}{3}\langle r_{1\Sigma^+}^2 \rangle-
\big [\frac{\kappa_{\Lambda^0}^2}{4m_{\Lambda^0}^2}-\frac{\kappa_{\Sigma^+}^2}{4m^2_{\Sigma^+}}\big ]=\\
\label{a44}
&=&\frac{2}{\pi^2\alpha}\int\limits_{\omega_{\Lambda^0}}^{\infty}
\frac{d\omega} {\omega}\big[\sigma_{tot}^{\gamma \Lambda^0\to
X}(\omega)- \sigma_{tot}^{\gamma \Sigma^+\to X}(\omega)\big ],
\end{eqnarray}

\begin{eqnarray}\nonumber
&&-\big [\frac{\kappa_{\Lambda^0}^2}{4m_{\Lambda^0}^2}-\frac{\kappa_{\Sigma^0}^2}{4m^2_{\Sigma^0}}\big ]=\\
\label{a45}
&=&\frac{2}{\pi^2\alpha}\int\limits_{\omega_{\Lambda^0}}^{\infty}
\frac{d\omega} {\omega}\big[\sigma_{tot}^{\gamma \Lambda^0\to
X}(\omega)- \sigma_{tot}^{\gamma \Sigma^0\to X}(\omega)\big ],
\end{eqnarray}

\begin{eqnarray}\nonumber
&&\frac{1}{3}\langle r_{1\Sigma^-}^2 \rangle-
\big [\frac{\kappa_{\Lambda^0}^2}{4m_{\Lambda^0}^2}-\frac{\kappa_{\Sigma^-}^2}{4m^2_{\Sigma^-}}\big ]=\\
\label{a46}
&=&\frac{2}{\pi^2\alpha}\int\limits_{\omega_{\Lambda^0}}^{\infty}
\frac{d\omega} {\omega}\big[\sigma_{tot}^{\gamma \Lambda^0\to
X}(\omega)- \sigma_{tot}^{\gamma \Sigma^-\to X}(\omega)\big ],
\end{eqnarray}

\begin{eqnarray}\nonumber
&&-\big [\frac{\kappa_{\Lambda^0}^2}{4m_{\Lambda^0}^2}-\frac{\kappa_{\Xi^0}^2}{4m^2_{\Xi^0}}\big ]=\\
\label{a47}
&=&\frac{2}{\pi^2\alpha}\int\limits_{\omega_{\Lambda^0}}^{\infty}
\frac{d\omega} {\omega}\big[\sigma_{tot}^{\gamma \Lambda^0\to
X}(\omega)- \sigma_{tot}^{\gamma \Xi^0\to X}(\omega)\big ],
\end{eqnarray}

\begin{eqnarray}\nonumber
&&\frac{1}{3}\langle r_{1\Xi^-}^2 \rangle-
\big [\frac{\kappa_{\Lambda^0}^2}{4m_{\Lambda^0}^2}-\frac{\kappa_{\Xi^-}^2}{4m^2_{\Xi^-}}\big ]=\\
\label{a48}
&=&\frac{2}{\pi^2\alpha}\int\limits_{\omega_{\Lambda^0}}^{\infty}
\frac{d\omega} {\omega}\big[\sigma_{tot}^{\gamma \Lambda^0\to
X}(\omega)- \sigma_{tot}^{\gamma \Xi^-\to X}(\omega)\big ],
\end{eqnarray}

\begin{eqnarray}\nonumber
&&\frac{1}{3}\langle r_{1\Sigma^+}^2 \rangle-
\big [\frac{\kappa_{\Sigma^+}^2}{4m_{\Sigma^+}^2}-\frac{\kappa_{\Xi^0}^2}{4m^2_{\Xi^0}}\big ]=\\
\label{a49}
&=&\frac{2}{\pi^2\alpha}\int\limits_{\omega_{\Sigma^+}}^{\infty}
\frac{d\omega} {\omega}\big[\sigma_{tot}^{\gamma \Sigma^+\to
X}(\omega)- \sigma_{tot}^{\gamma \Xi^0\to X}(\omega)\big ],
\end{eqnarray}

\begin{eqnarray}\nonumber
&&\frac{1}{3}\big [\langle r_{1\Sigma^+}^2 \rangle+\langle
r_{1\Xi^-}^2 \rangle\big ]-
\big [\frac{\kappa_{\Sigma^+}^2}{4m_{\Sigma^+}^2}-\frac{\kappa_{\Xi^-}^2}{4m^2_{\Xi^-}}\big ]=\\
\label{a50}
&=&\frac{2}{\pi^2\alpha}\int\limits_{\omega_{\Sigma^+}}^{\infty}
\frac{d\omega} {\omega}\big[\sigma_{tot}^{\gamma \Sigma^+\to
X}(\omega)- \sigma_{tot}^{\gamma \Xi^-\to X}(\omega)\big ],
\end{eqnarray}

\begin{eqnarray}\nonumber
&&-\big [\frac{\kappa_{\Sigma^0}^2}{4m_{\Sigma^0}^2}-\frac{\kappa_{\Xi^0}^2}{4m^2_{\Xi^0}}\big ]=\\
\label{a51}
&=&\frac{2}{\pi^2\alpha}\int\limits_{\omega_{\Sigma^0}}^{\infty}
\frac{d\omega} {\omega}\big[\sigma_{tot}^{\gamma \Sigma^0\to
X}(\omega)- \sigma_{tot}^{\gamma \Xi^0\to X}(\omega)\big ],
\end{eqnarray}

\begin{eqnarray}\nonumber
&&\frac{1}{3}\langle r_{1\Xi^-}^2 \rangle-
\big [\frac{\kappa_{\Sigma^0}^2}{4m_{\Sigma^0}^2}-\frac{\kappa_{\Xi^-}^2}{4m^2_{\Xi^-}}\big ]=\\
\label{a52}
&=&\frac{2}{\pi^2\alpha}\int\limits_{\omega_{\Sigma^0}}^{\infty}
\frac{d\omega} {\omega}\big[\sigma_{tot}^{\gamma \Sigma^0\to
X}(\omega)- \sigma_{tot}^{\gamma \Xi^-\to X}(\omega)\big ],
\end{eqnarray}

\begin{eqnarray}\nonumber
&&-\frac{1}{3}\langle r_{1\Sigma^-}^2 \rangle-
\big [\frac{\kappa_{\Sigma^-}^2}{4m_{\Sigma^-}^2}-\frac{\kappa_{\Xi^0}^2}{4m^2_{\Xi^0}}\big ]=\\
\label{a53}
&=&\frac{2}{\pi^2\alpha}\int\limits_{\omega_{\Sigma^-}}^{\infty}
\frac{d\omega} {\omega}\big[\sigma_{tot}^{\gamma \Sigma^-\to
X}(\omega)- \sigma_{tot}^{\gamma \Xi^0\to X}(\omega)\big ],
\end{eqnarray}

\begin{eqnarray}\nonumber
&&\frac{1}{3}\big [-\langle r_{1\Sigma^-}^2 \rangle+\langle
r_{1\Xi^-}^2 \rangle\big ]-
\big [\frac{\kappa_{\Sigma^-}^2}{4m_{\Sigma^-}^2}-\frac{\kappa_{\Xi^-}^2}{4m^2_{\Xi^-}}\big ]=\\
\label{a54}
&=&\frac{2}{\pi^2\alpha}\int\limits_{\omega_{\Sigma^-}}^{\infty}
\frac{d\omega} {\omega}\big[\sigma_{tot}^{\gamma \Sigma^-\to
X}(\omega)- \sigma_{tot}^{\gamma \Xi^-\to X}(\omega)\big ].
\end{eqnarray}

\begin{table*}[t]
\caption{}
\begin{ruledtabular}
\begin{tabular}{c c c c c c c }\hline
B &  $I_B[mb]$ & $m_B [Gev]$ & $\kappa_B [\mu_N]$&$\langle r_{EB}^2
\rangle [fm^2]$&$3\kappa_B/2m_B^2 [fm^2]$&$\langle r_{1B}^2 \rangle
[fm^2]$\\  \hline
$p$  &0.9125         & 0.93827  & 1.7928 & 0.717& 0.119 & 0.598      \\
$n$ & 0.9100    & 0.93957    &-1.9130    &-0.113  &-0.127 & -0.240     \\
$\Lambda^0$  & 0.6454 &  1.11568   & -0.6130 & 0.110 & -0.029 & 0.081   \\
$\Sigma^+$ & 0.5679 & 1.18937 & 1.4580 & 0.600 & 0.060 & 0.660 \\
$\Sigma^0$& 0.5648 & 1.19264 & 0.6490    & -0.030  & 0.027 & -0.003     \\
$\Sigma^-$ & 0.5602   & 1.19745   & -0.1600   & 0.670 & -0.007 & 0.663   \\
$\Xi^0$  & 0.4647    & 1.31483 & -1.2500 & 0.130 & -0.042    & 0.088     \\
$\Xi^-$ &0.4601  & 1.32131 & 0.3493 & 0.490  & 0.012  & 0.502 \\
\end{tabular}
\end{ruledtabular}
\end{table*}

In order to evaluate the left hand sides of the derived sum rules
and to draw out some phenomenological consequences, one needs the
reliable values of Dirac baryon mean square radii $\langle
r_{1B}^2 \rangle$ and baryon anomalous magnetic moments
$\kappa_B$.

The latter are known (besides $\Sigma^0$, which is found from the
well known relation $\kappa_{\Sigma^+}+
\kappa_{\Sigma^-}$=$2\kappa_{\Sigma^0}$) experimentally (see the
third column in Table I), however, to calculate $\langle r_{1B}^2
\rangle$ by means of the difference of the baryon electric mean
square radius $\langle r_{EB}^2 \rangle$ and Foldy term, well
known for all ground state octet baryons from the experimental
information on the magnetic moments given by Review of Particle
Physics \cite{Rev}
\begin{equation}
\langle r_{1B}^2 \rangle =\langle r_{EB}^2
\rangle-\frac{3\kappa_B}{2 m_B^2}, \label{a56}
\end{equation}
we are in need of the reliable values of $\langle r_{EB}^2
\rangle$.

They are known experimentally only for the proton, neutron and
$\Sigma^-$-hyperon.

Fortunately there are recent results \cite{Kub} to fourth order in
relativistic baryon chiral perturbation theory (giving predictions
for the $\Sigma^-$ charge radius and the $\Lambda$-$\Sigma^0$
transition moment in excellent agreement with the available
experimental information), which solve our problem completely.

All necessary information is collected in Table I, where also
numerical values of corresponding $\langle r^2_{1B} \rangle$ are
presented.

 Calculating the left-hand side of all sum rules one finds
{ \begin{widetext}
\begin{equation}
\label{a57} \frac{2}{\pi^2\alpha}\int\limits_{\omega_{p}}^{\infty}
\frac{d\omega} {\omega}\big[\sigma_{tot}^{\gamma p\to X}(\omega)-
\sigma_{tot}^{\gamma n\to X}(\omega)\big ]=2.0415
  {\textrm mb},\quad \Rightarrow \quad \sigma_{tot}^{\gamma p \to X}(\omega)> \sigma_{tot}^{\gamma
n\to X}(\omega)
\end{equation}

\begin{equation}
\label{a58}
\frac{2}{\pi^2\alpha}\int\limits_{\omega_{\Sigma^+}}^{\infty}
\frac{d\omega} {\omega}\big[\sigma_{tot}^{\gamma \Sigma^+\to
X}(\omega)- \sigma_{tot}^{\gamma \Sigma^0\to X}(\omega)\big
]=2.0825
  {\textrm mb},\quad \Rightarrow \quad \sigma_{tot}^{\gamma \Sigma^+\to X}(\omega)>
\sigma_{tot}^{\gamma \Sigma^0\to X}(\omega)
\end{equation}

\begin{equation}
\label{a59}
\frac{2}{\pi^2\alpha}\int\limits_{\omega_{\Sigma^+}}^{\infty}
\frac{d\omega} {\omega}\big[\sigma_{tot}^{\gamma \Sigma^+\to
X}(\omega)- \sigma_{tot}^{\gamma \Sigma^-\to X}(\omega) \big
]=4.2654  {\textrm mb}, \quad \Rightarrow\quad
\sigma_{tot}^{\gamma \Sigma^+\to X}(\omega)> \sigma_{tot}^{\gamma
\Sigma^-\to X}(\omega)
\end{equation}

\begin{equation}
\label{a60}
\frac{2}{\pi^2\alpha}\int\limits_{\omega_{\Sigma^0}}^{\infty}
\frac{d\omega} {\omega}\big[\sigma_{tot}^{\gamma \Sigma^0\to
X}(\omega)- \sigma_{tot}^{\gamma \Sigma^-\to X}(\omega)\big ]=
2.1829  {\textrm mb}, \quad \Rightarrow\quad \sigma_{tot}^{\gamma
\Sigma^0\to X}(\omega)> \sigma_{tot}^{\gamma \Sigma^-\to
X}(\omega)
\end{equation}

\begin{equation}
\label{a61}
\frac{2}{\pi^2\alpha}\int\limits_{\omega_{\Xi^0}}^{\infty}
\frac{d\omega} {\omega}\big[\sigma_{tot}^{\gamma \Xi^0\to
X}(\omega)- \sigma_{tot}^{\gamma \Xi^-\to X}(\omega)\big ]=1.5921
    {\textrm mb}, \quad \Rightarrow\quad \sigma_{tot}^{\gamma
\Xi^0\to X}(\omega)> \sigma_{tot}^{\gamma \Xi^-\to X}(\omega)
\end{equation}

\begin{equation}
\label{a62} \frac{2}{\pi^2\alpha}\int\limits_{\omega_{p}}^{\infty}
\frac{d\omega} {\omega}\big[\sigma_{tot}^{\gamma p\to X}(\omega)-
\sigma_{tot}^{\gamma \Lambda^0\to X}(\omega)\big ]=1.6673 {\textrm
mb},\quad \Rightarrow\quad \sigma_{tot}^{\gamma p\to X}(\omega)>
\sigma_{tot}^{\gamma \Lambda^0\to X}(\omega)
\end{equation}

\begin{equation}
\label{a63} \frac{2}{\pi^2\alpha}\int\limits_{\omega_{p}}^{\infty}
\frac{d\omega} {\omega}\big[\sigma_{tot}^{\gamma p \to X}(\omega)-
\sigma_{tot}^{\gamma \Sigma^+\to X}(\omega)\big ]=-0.4158 {\textrm
mb},\quad \Rightarrow\quad \sigma_{tot}^{\gamma p \to
X}(\omega)<\sigma_{tot}^{\gamma \Sigma^+\to X}(\omega)
\end{equation}

\begin{equation}
\label{a64} \frac{2}{\pi^2\alpha}\int\limits_{\omega_{p}}^{\infty}
\frac{d\omega} {\omega}\big[\sigma_{tot}^{\gamma p\to X}(\omega)-
\sigma_{tot}^{\gamma \Sigma^0\to X}(\omega)\big ]=1.6667  {\textrm
mb},\quad \Rightarrow\quad \sigma_{tot}^{\gamma p\to X}(\omega)>
\sigma_{tot}^{\gamma \Sigma^0\to X}(\omega)
\end{equation}

\begin{equation}
\label{a65} \frac{2}{\pi^2\alpha}\int\limits_{\omega_{p}}^{\infty}
\frac{d\omega} {\omega}\big[\sigma_{tot}^{\gamma p \to X}(\omega)-
\sigma_{tot}^{\gamma \Sigma^-\to X}(\omega)\big ]= 3.8496 {\textrm
mb},\quad \Rightarrow\quad \sigma_{tot}^{\gamma p \to X}(\omega)>
\sigma_{tot}^{\gamma \Sigma^-\to X}(\omega)
\end{equation}

\begin{equation}
\label{a66} \frac{2}{\pi^2\alpha}\int\limits_{\omega_{p}}^{\infty}
\frac{d\omega} {\omega}\big[\sigma_{tot}^{\gamma p\to X}(\omega)-
\sigma_{tot}^{\gamma \Xi^0\to X}(\omega)\big ]=1.7259  {\textrm
mb},\quad  \Rightarrow \quad \sigma_{tot}^{\gamma p\to X}(\omega)>
\sigma_{tot}^{\gamma \Xi^0\to X}(\omega)
\end{equation}

\begin{equation}
\label{a67} \frac{2}{\pi^2\alpha}\int\limits_{\omega_{p}}^{\infty}
\frac{d\omega} {\omega}\big[\sigma_{tot}^{\gamma p \to X}(\omega)-
\sigma_{tot}^{\gamma \Xi^-\to X}(\omega)\big ]=3.3180   {\textrm
mb},\quad \Rightarrow\quad \sigma_{tot}^{\gamma p \to X}(\omega)>
\sigma_{tot}^{\gamma \Xi^-\to X}(\omega)
\end{equation}

\begin{equation}
\label{a68} \frac{2}{\pi^2\alpha}\int\limits_{\omega_{n}}^{\infty}
\frac{d\omega} {\omega}\big[\sigma_{tot}^{\gamma n\to X}(\omega)-
\sigma_{tot}^{\gamma \Lambda^0\to X}(\omega)\big ]=-0.3260
{\textrm mb},\quad \Rightarrow\quad \sigma_{tot}^{\gamma n\to
X}(\omega)< \sigma_{tot}^{\gamma \Lambda^0\to X}(\omega)
\end{equation}

\begin{equation}
\label{a69} \frac{2}{\pi^2\alpha}\int\limits_{\omega_{n}}^{\infty}
\frac{d\omega} {\omega}\big[\sigma_{tot}^{\gamma n\to X}(\omega)-
\sigma_{tot}^{\gamma \Sigma^+\to X}(\omega)\big ]=-2.4573 {\textrm
mb},\quad \Rightarrow\quad \sigma_{tot}^{\gamma n\to X}(\omega)<
\sigma_{tot}^{\gamma \Sigma^+\to X}(\omega)
\end{equation}

\begin{equation}
\label{a70} \frac{2}{\pi^2\alpha}\int\limits_{\omega_{n}}^{\infty}
\frac{d\omega} {\omega}\big[\sigma_{tot}^{\gamma n\to X}(\omega)-
\sigma_{tot}^{\gamma \Sigma^0\to X}(\omega)\big ]=-0.3747 {\textrm
mb},\quad \Rightarrow\quad \sigma_{tot}^{\gamma n\to X}(\omega)<
\sigma_{tot}^{\gamma \Sigma^0\to X}(\omega)
\end{equation}

\begin{equation}
\label{a71} \frac{2}{\pi^2\alpha}\int\limits_{\omega_{n}}^{\infty}
\frac{d\omega} {\omega}\big[\sigma_{tot}^{\gamma n\to X}(\omega)-
\sigma_{tot}^{\gamma \Sigma^-\to X}(\omega)\big ]= 1.8082 {\textrm
mb},\quad \Rightarrow\quad \sigma_{tot}^{\gamma n\to X}(\omega)>
\sigma_{tot}^{\gamma \Sigma^-\to X}(\omega)
\end{equation}

\begin{equation}
\label{a72} \frac{2}{\pi^2\alpha}\int\limits_{\omega_{n}}^{\infty}
\frac{d\omega} {\omega}\big[\sigma_{tot}^{\gamma n\to X}(\omega)-
\sigma_{tot}^{\gamma \Xi^0\to X}(\omega)\big ]= -0.3156  {\textrm
mb},\quad \Rightarrow\quad \sigma_{tot}^{\gamma n\to X}(\omega)<
\sigma_{tot}^{\gamma \Xi^0\to X}(\omega)
\end{equation}

\begin{equation}
\label{a73} \frac{2}{\pi^2\alpha}\int\limits_{\omega_{n}}^{\infty}
\frac{d\omega} {\omega}\big[\sigma_{tot}^{\gamma n\to X}(\omega)-
\sigma_{tot}^{\gamma \Xi^-\to X}(\omega)\big ]=1.2766  {\textrm
mb},\quad \Rightarrow\quad \sigma_{tot}^{\gamma n\to X}(\omega)>
\sigma_{tot}^{\gamma \Xi^-\to X}(\omega)
\end{equation}

\begin{equation}
\label{a74}
\frac{2}{\pi^2\alpha}\int\limits_{\omega_{\Lambda^0}}^{\infty}
\frac{d\omega} {\omega}\big[\sigma_{tot}^{\gamma \Lambda^0\to
X}(\omega)- \sigma_{tot}^{\gamma \Sigma^+\to X}(\omega)\big ]=
-2.0831 {\textrm mb},\quad \Rightarrow\quad \sigma_{tot}^{\gamma
\Lambda^0\to X}(\omega)< \sigma_{tot}^{\gamma \Sigma^+\to
X}(\omega)
\end{equation}

\begin{equation}
\label{a75}
\frac{2}{\pi^2\alpha}\int\limits_{\omega_{\Lambda^0}}^{\infty}
\frac{d\omega} {\omega}\big[\sigma_{tot}^{\gamma \Lambda^0\to
X}(\omega)- \sigma_{tot}^{\gamma \Sigma^0\to X}(\omega)\big ]=
-0.0006   {\textrm mb},\quad \Rightarrow\quad \sigma_{tot}^{\gamma
\Lambda^0\to X}(\omega)\approx \sigma_{tot}^{\gamma \Sigma^0\to
X}(\omega)
\end{equation}

\begin{equation}
\label{a76}
\frac{2}{\pi^2\alpha}\int\limits_{\omega_{\Lambda^0}}^{\infty}
\frac{d\omega} {\omega}\big[\sigma_{tot}^{\gamma \Lambda^0\to
X}(\omega)- \sigma_{tot}^{\gamma \Sigma^-\to X}(\omega)\big
]=2.1823
  {\textrm{mb}},\quad \Rightarrow\quad \sigma_{tot}^{\gamma
\Lambda^0\to X}(\omega)> \sigma_{tot}^{\gamma \Sigma^-\to X}(\omega)
\end{equation}

\begin{equation}
\label{a77}
\frac{2}{\pi^2\alpha}\int\limits_{\omega_{\Lambda^0}}^{\infty}
\frac{d\omega} {\omega}\big[\sigma_{tot}^{\gamma \Lambda^0\to
X}(\omega)- \sigma_{tot}^{\gamma \Xi^0\to X}(\omega)\big ]=0.0586
{\textrm{mb}},\quad \Rightarrow\quad \sigma_{tot}^{\gamma
\Lambda^0\to X}(\omega)> \sigma_{tot}^{\gamma \Xi^0\to X}(\omega)
\end{equation}

\begin{equation}
\label{a78}
\frac{2}{\pi^2\alpha}\int\limits_{\omega_{\Lambda^0}}^{\infty}
\frac{d\omega} {\omega}\big[\sigma_{tot}^{\gamma \Lambda^0\to
X}(\omega)- \sigma_{tot}^{\gamma \Xi^-\to X}(\omega)\big ]=2.1823
{\textrm{mb}} ,\quad \Rightarrow\quad \sigma_{tot}^{\gamma
\Lambda^0\to X}(\omega)> \sigma_{tot}^{\gamma \Xi^-\to X}(\omega)
\end{equation}

\begin{equation}
\label{a79}
\frac{2}{\pi^2\alpha}\int\limits_{\omega_{\Sigma^+}}^{\infty}
\frac{d\omega} {\omega}\big[\sigma_{tot}^{\gamma \Sigma^+\to
X}(\omega)- \sigma_{tot}^{\gamma \Xi^0\to X}(\omega)\big ]=2.1417
\textrm{mb},\quad \Rightarrow\quad \sigma_{tot}^{\gamma
\Sigma^+\to X}(\omega)> \sigma_{tot}^{\gamma \Xi^0\to X}(\omega)
\end{equation}

\begin{equation}
\label{a80}
\frac{2}{\pi^2\alpha}\int\limits_{\omega_{\Sigma^+}}^{\infty}
\frac{d\omega} {\omega}\big[\sigma_{tot}^{\gamma \Sigma^+\to
X}(\omega)- \sigma_{tot}^{\gamma \Xi^-\to X}(\omega)\big ]=3.7338
\textrm{mb},\quad \Rightarrow\quad \sigma_{tot}^{\gamma
\Sigma^+\to X}(\omega)> \sigma_{tot}^{\gamma \Xi^-\to X}(\omega)
\end{equation}

\begin{equation}
\label{a81}
\frac{2}{\pi^2\alpha}\int\limits_{\omega_{\Sigma^0}}^{\infty}
\frac{d\omega} {\omega}\big[\sigma_{tot}^{\gamma \Sigma^0\to
X}(\omega)- \sigma_{tot}^{\gamma \Xi^0\to X}(\omega)\big ]=0.1168
\textrm{mb},\quad \Rightarrow\quad \sigma_{tot}^{\gamma
\Sigma^0\to X}(\omega)> \sigma_{tot}^{\gamma \Xi^0\to X}(\omega)
\end{equation}

\begin{equation}
\label{a82}
\frac{2}{\pi^2\alpha}\int\limits_{\omega_{\Sigma^0}}^{\infty}
\frac{d\omega} {\omega}\big[\sigma_{tot}^{\gamma \Sigma^0\to
X}(\omega)- \sigma_{tot}^{\gamma \Xi^-\to X}(\omega)\big ]=1.5732
\textrm{mb},\quad \Rightarrow\quad \sigma_{tot}^{\gamma
\Sigma^0\to X}(\omega)> \sigma_{tot}^{\gamma \Xi^\to X}(\omega)
\end{equation}

\begin{equation}
\label{a83}
\frac{2}{\pi^2\alpha}\int\limits_{\omega_{\Sigma^-}}^{\infty}
\frac{d\omega} {\omega}\big[\sigma_{tot}^{\gamma \Sigma^-\to
X}(\omega)- \sigma_{tot}^{\gamma \Xi^0\to X}(\omega)\big ]=-2.1238
\textrm{mb},\quad \Rightarrow\quad \sigma_{tot}^{\gamma
\Sigma^-\to X}(\omega)< \sigma_{tot}^{\gamma \Xi^0\to X}(\omega)
\end{equation}

\begin{equation}
\label{a84}
\frac{2}{\pi^2\alpha}\int\limits_{\omega_{\Sigma^-}}^{\infty}
\frac{d\omega} {\omega}\big[\sigma_{tot}^{\gamma \Sigma^-\to
X}(\omega)- \sigma_{tot}^{\gamma \Xi^-\to X}(\omega)\big ]=-0.5316
\textrm{mb},\quad \Rightarrow\quad \sigma_{tot}^{\gamma
\Sigma^-\to X}(\omega)< \sigma_{tot}^{\gamma \Xi^-\to X}(\omega),
\end{equation}
\end{widetext}}

from where one gets the following chain of inequalities
{\begin{widetext}
\begin{eqnarray*}
\sigma_{tot}^{\gamma \Sigma^+\to X}(\omega)> \sigma_{tot}^{\gamma
p\to X}(\omega)> \sigma_{tot}^{\gamma \Lambda^0\to X}(\omega)
\approx   \sigma_{tot}^{\gamma \Sigma^0\to X}(\omega)>\\
\sigma_{tot}^{\gamma \Xi^0\to X}(\omega)>
 \sigma_{tot}^{\gamma n\to X}(\omega)>
\sigma_{tot}^{\gamma \Xi^-\to X}(\omega)>\sigma_{tot}^{\gamma
\Sigma^-\to X}(\omega)
\end{eqnarray*}
\end{widetext}}
for total cross-sections of hadron photo-production on ground
state $1/2^+$ octet baryons to be valid in average for finite
values of $\omega$.

\section{CONCLUSIONS}

Considering the very high energy peripheral electron-hadron
scattering with a production of a hadronic state $X$ moving
closely to the direction of initial hadron, then exploiting
analytic properties of the forward retarded Compton scattering
amplitude on the same hadron, for the case of small transferred
momenta, new meson and baryon sum rules were derived. Evaluating
the left-hand sides of the sum rules, the chains of inequalities
for total cross-sections of hadron photo-production on
pseudoscalar mesons and $1/2^+$ octet baryons have been found.

\end{document}